\documentclass[dvips]{acta}
\usepackage{supertabular,lscape,epsfig}
\usepackage{amssymb}
\usepackage{amsmath}
\SetPages{1}{9}
\SetVol{72}{2022}

\begin{document}

\begin{Titlepage}

\Title { On the Orbital Period and Models of V Sge }

\Author {J.~~S m a k}
{N. Copernicus Astronomical Center, Polish Academy of Sciences,\\
Bartycka 18, 00-716 Warsaw, Poland\\
e-mail: jis@camk.edu.pl }

\Received{  }

\end{Titlepage}

\begin {flushright} 
To George Preston on his 92nd Birthday
\end {flushright}

\Abstract { 
The orbital period of V Sge is decreasing at a rate which increased from 
$dP/dt=-(4.11\pm 0.33)\times 10^{-10}$ in 1962 to $-(5.44\pm 0.61)\times 10^{-10}$ 
in 2022. This implies that the mass trahsfer from the secondary component 
is accelerating.  

From the evidence based on the orbital period variations, combined with 
estimates of the mass loss from the system based on radio observations, 
it follows that 
(1) the mass transfer rate from the secondary component is larger than 
$\dot M_2=-5\times 10^{-6}M{\odot }/yr$, possibly as large as 
$\dot M_2=-2.5\times 10^{-5}M{\odot }/yr$, 
and (2) the mass loss rate from the primary component is  
$\dot M_1=-4\times 10^{-7}M{\odot }/yr$ or larger. 

Close similarity of V Sge to binary Wolf-Rayet stars supports the model with 
primary component being a hot, evolved star loosing its mass.  
Several arguments are presented which exclude the alternative model with 
primary component being a white dwarf with an accretion disk. 
} 

{binaries: close -- stars: winds, outflows -- X-rays: stars -- stars: 
 mass loss -- stars: individual: V Sge }

\section { Introduction } 

Sixty years ago Herbig {\it et al.} (1965) discovered V Sge to be the double-line  
spectroscopic and eclipsing binary. Using results of their extensive spectroscopic 
and photometric observations they presented a model of an interacting binary system 
with two hot components and an extended gaseous envelope. 
Shortly afterwords it was found that the orbital period is decreasing (Smak 1967). 
In the following years V Sge was observed by numerous authors in different spectral 
regions and the original model by Herbig {\it et al.} was significantly improved 
and/or modified (see, e.g., Smak, Belczy\'nski and Zo{\l}a 2001, for an earlier 
review). 

The purpose of the present paper is twofold: (1) to re-discuss the variations 
of the orbital period (Sections 2 and 3) and (2) to critically review the two 
competing models of V Sge (Section 4). 

\section { The (O-C) Diagram } 

Moments of minima of V Sge observed prior to the year 1997 were listed 
by Lockley {\it et al.} (1997) in their Table 2. Those observed after that year 
have been collected from the literature and are listed in Table 1. 

\begin{table}[h!]
{\parskip=0truept
\baselineskip=0pt {
\medskip
\centerline{Table 1}
\medskip
\centerline{ Minima of V Sge }
\medskip
$$\offinterlineskip \tabskip=0pt
\vbox {\halign {\strut
\vrule width 0.5truemm #&	
\hfil\quad#\quad\hfil&            
\vrule#&			
\hfil\quad#\quad\hfil&            
\vrule#&			
\hfil\quad#\quad\hfil&            
\vrule#&			
\hfil\quad#\quad\hfil&            
\vrule#&			
\hfil\quad#\quad\hfil&            
\vrule#&			
\hfil\quad#\quad\hfil&            
\vrule width 0.5 truemm # \cr	
\noalign {\hrule height 0.5truemm}
&  JDhel.    &&     &&  JDhel.    &&     &&  JDhel.    &&     &\cr
&  2400000+  && ref &&  2400000+  && ref &&  2400000+  && ref &\cr
\noalign {\hrule height 0.5truemm}
& 50969.4725 && (1) && 53900.3619 && (5) && 57220.4758 && (9) &\cr
& 51781.3817 && (2) && 53902.4117 && (5) && 57235.3855 && (10) &\cr
& 52817.4834 && (3) && 53940.4648 && (5) && 57237.4388 && (10) &\cr
& 53217.5102 && (4) && 53972.3476 && (5) && 57272.3977 && (10) &\cr
& 53233.4622 && (4) && 53975.4324 && (5) && 57273.4345 && (10) &\cr
& 53246.3112 && (4) && 53991.3636 && (5) && 57589.6577 && (10) &\cr
& 53265.3322 && (4) && 53992.4066 && (5) && 57590.6892 && (10) &\cr
& 53266.3649 && (4) && 53993.4375 && (5) && 57615.3703 && (10) &\cr
& 53267.3896 && (4) && 53993.4427 && (5) && 57975.8248 && (10) &\cr
& 53282.3009 && (4) && 54007.3107 && (5) && 57983.5202 && (10) &\cr
& 53283.3321 && (4) && 54023.2561 && (5) && 57988.6675 && (10) &\cr
& 53284.3587 && (4) && 54024.2779 && (5) && 57989.6933 && (10) &\cr
& 53285.3871 && (4) && 54026.3407 && (5) && 57990.7243 && (10) &\cr 
& 53579.5023 && (5) && 54388.3306 && (6) && 58006.6587 && (10) &\cr 
& 53580.5293 && (5) && 55028.4893 && (7) && 58007.6907 && (10) &\cr
& 53581.5574 && (5) && 55097.3965 && (7) && 59067.4345 && (10) &\cr
& 53596.4716 && (5) && 56891.3994 && (8) && 59068.4595 && (10) &\cr
& 53615.5040 && (5) && 57219.4382 && (9) &&&&&\cr
\noalign {\hrule height 0.5truemm}
}}$$
Notes to Table 1: (1) Ogloza {\it et al.} (2000). (2) Zejda (2002). 
(3) H{\"o}bscher (2005). (4) Pribulla {\it et al.} (2005). (5) Parimucha (2007). 
(6) H{\"o}bscher {\it et al.} (2008). (7) H{\"o}bscher {\it et al.} (2010).  
(8) H{\"o}bscher and Lehmann (2015). (9) H{\"o}bscher (2016). 
(10) determined from AAVSO light curves.
}}
\end{table}

The values of (O-C), calculated from the original elements given by 
Herbig {\it et al.} (1965) 

\beq
Pri.Min~=~JDhel.~2437889.9154~+~0.514195\times E~,
\eeq 

\noindent
are plotted in Fig.1. 

\begin{figure}[htb]
\vskip -20truemm
\epsfysize=12.0cm 
\hspace{2.0cm}
\epsfbox{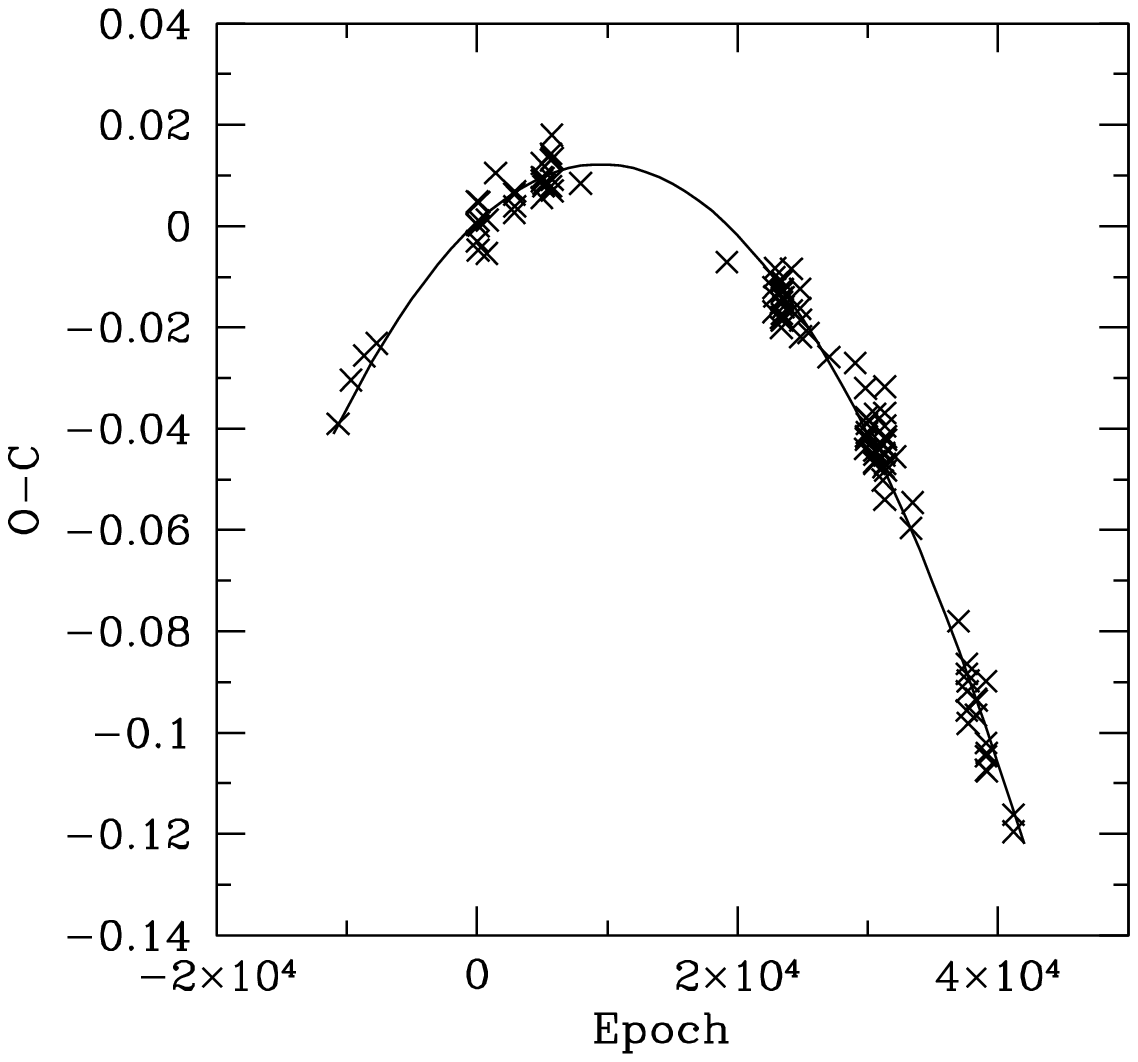} 
\vskip -25truemm
\FigCap { The (O-C) diagram calculated from the original elements given by 
Herbig {\it et al.} (1965). The line represents Eq.(4). }
\end{figure}

Fitting a parabola to the points in Fig.1 we get  

\beq
(O-C)~=~0.0008(8)~+~0.00000241(10)E~-~1.270(26)\times 10^{-10}E^2~,
\eeq 

\noindent
and the corresponding period decrase

\beq
{{dP}\over{dt}}~=~(-4.94\pm 0.10)\times 10^{-10}~. 
\eeq 

Fitting -- for the first time(!) -- a third degree polynomial gives 

\bdm
(O-C)~=~0.0001(8)~+~0.00000218(13)E~-~1.057(86)\times 10^{-10}E^2~
\edm

\vskip -5truemm

\beq
   ~~~~~~    -~0.41(16)\times 10^{-15}E^3~, 
\eeq

\noindent 
and the corresponding period derivatives 

\beq
\left({{dP}\over{dt}}\right)_{1962}~=~(-4.11\pm 0.33)\times 10^{-10}~,~~~ 
{{d^2P}\over{dt^2}}~=~(-6.1\pm 2.3)\times 10^{-15}~[d^{-1}]~. 
\eeq 

\noindent
The value of $dP/dt$ given above refers to $E=0$, i.e. to the year 1962. 
For completeness we give its value refering to the year 2022

\beq
\left({{dP}\over{dt}}\right)_{2022}~=~(-5.44\pm 0.61)\times 10^{-10}~. 
\eeq 

\noindent
This shows that the rate at which the orbital period is decreasing is 
increasing and implies that the process of mass trahsfer from the secondary 
component is accelerating.

\section { The Mass Transfer and Mass Loss Rates } 

In what follows we use system parameters obtained earlier by us (Smak {\it et al.} 
2001). And, in particular, we refer to the {\it mean brightness} of V Sge 
("MS-low" in our notation). 
To begin with, using formulae given by Paczy\'nski (1971, Eqs.10 and 11), 
we obtain the following estimates of the thermal time scale and the mass transfer rate 
from the secondary component: 

\beq
\tau_{K-H,2}=1.3\times 10^5~yrs, ~~~~ {{dM_2}\over{dt}}=
   \dot M_2=-~2.6\times 10^{-5}M{\odot}/yr~.
\eeq 

\noindent
Turning to the primary component we find (see Appendix) that even in the faintest state 
("HPSP-faint 1") its bolometric luminosity is larger than the corresponding Eddington 
luminosity. This implies substantial mass loss from this component. 

We also have an independent estimate of the mass loss from the system based on radio 
observations made by Lockley {\it et al.} (1997,1999).   
From the flux at $\lambda=3.6$cm, measured during two seasons: $S=0.07$ and 0.45mJy,  
assuming distance $d=2.75$kpc, they obtained 
$\dot M_{1+2}=-(0.7-2.8)\times 10^{-5}M{\odot }/yr$. 
Using our distance $d=4$kps we get $\dot M_{1+2}=-(1.2-4.9)\times 10^{-5}M{\odot}/yr$ 
or $\log \dot M_{1+2}=-(4.9-4.3)$; we adopt

\beq
\log \dot M_{1+2}~=~-~4.6\pm 0.3~.
\eeq 

We now turn to the evidence from the observed period variations. 
In general they can be described by 

\beq
{{d\ln P}\over{dt}}~=~3{{d\ln J}\over{dt}}~+~{{d\ln (M_1+M_2)}\over{dt}}~
         -~3{{d\ln M_1}\over{dt}}~-~3{{d\ln M_2}\over{dt}}~,
\eeq 

\noindent
where $J$ is the total angular momentum of the system. 

Our goal is to obtain a relation between the two mass transfer/mass loss rates: 
$dM_1/dt=\dot M_1$ and $dM_2/dt=\dot M_2$. 
We begin with the angular momentum which is removed from the system together with 
the escaping material and consider two cases: 

{\it Case 1}. The escaping material has the specific angular momentum identical 
with that of component 1 

\beq
j=j_1=q<j> ~~~ {\rm and} ~~~ {{d\ln J}\over{dt}}=q~{{d\ln (M_1+M_2)}\over{dt}}~. 
\eeq 

\noindent
The resulting $\dot M_1-\dot M_2$ relation becomes: 

\beq
{{d\ln P}\over{dt}}~=~-~{2\over{1+q}} {{d\ln M_1}\over{dt}}~+~
      {{3q^2-2q-3}\over {1+q}} {{d\ln M_2}\over{dt}}~.
\eeq 

{\it Case 2}. The escaping material has the specific angular momentum identical 
with the mean angular momentum of the system 

\beq
j=<j>, ~~~ {\rm and} ~~~ {{d\ln J}\over{dt}}~=~{{d\ln (M_1+M_2)}\over{dt}}~. 
\eeq 

\noindent
The resulting $\dot M_1-\dot M_2$ relation becomes: 

\beq
{{d\ln P}\over{dt}}~=~-~{{3q-1}\over{1+q}} {{d\ln M_1}\over{dt}}~+~
      {{q-3}\over {1+q}} {{d\ln M_2}\over{dt}}~.
\eeq 

\begin{figure}[htb]
\vskip -15truemm
\epsfysize=8.0cm 
\hspace{2.5cm}
\epsfbox{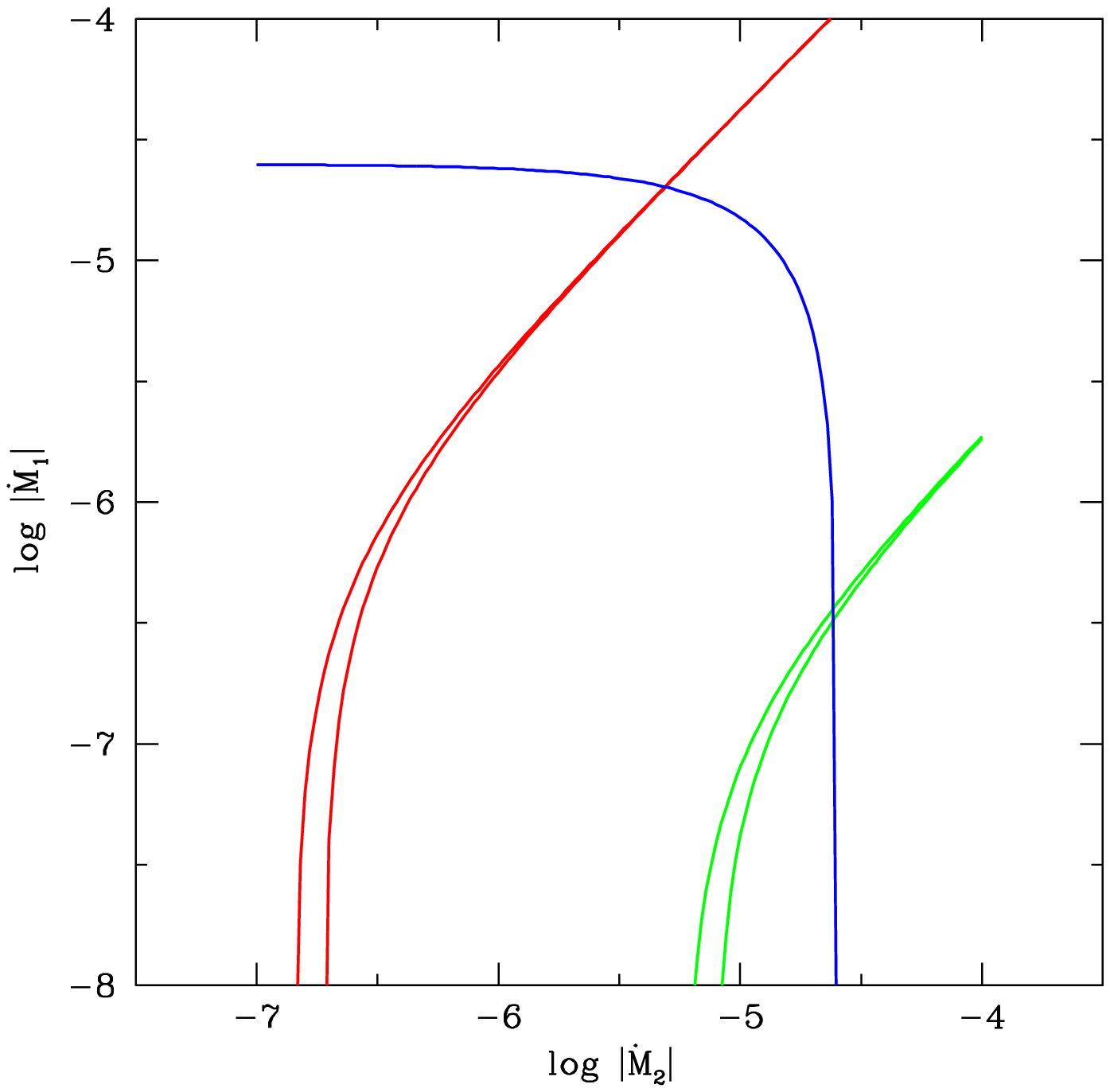} 
\vskip -5truemm
\FigCap { The $\dot M_1-\dot M_2$ diagram. 
The lines represent Eq.(11) ({\it green}), Eq.(13) ({\it red}), and 
Eq.(8) ({\it blue}). Double lines correspond to the the two values of $dP/dt$. }
\end{figure}

The results are presented in Fig.2. From the intersections of the two 
$\dot M_1-\dot M_2$ relations with the $\dot M_{1+2}$ line we get crude estimates: 

\beq
\dot M_1=-2\times 10^{-5}~M{\odot}/yr~~~{\rm and}~~~\dot M_2=-5\times 10^{-6}
    ~M{\odot}/yr~,
\eeq 

\noindent
for Case 1 and

\beq
\dot M_1=-4\times 10^{-7}~M{\odot}/yr~~~{\rm and}~~~\dot M_2=-2.5\times 10^{-5}
    ~M{\odot}/yr~,
\eeq 

\noindent
for Case 2. Worth noting is that $\dot M_2$ obtained for this case agrees with 
the value predicted above (Eq.7). This may indicate that Case 2 -- better than 
Case 1 -- describes the loss of angular momentum from the system. 
If so, we can only conclude that 
(1) the mass transfer rate from the secondary component is larger than 
$\dot M_2=-5\times 10^{-6}~M{\odot}/yr$, possibly as large as 
$\dot M_2=-2.5\times 10^{-5}~M{\odot}/yr$, and (2) the mass loss rate from 
the primary component is larger than $\dot M_1=-4\times 10^{-7}~M{\odot}/yr$.

\section { The Two Competing Models } 

Two models were proposed to explain the observed properties and peculiarities 
of V Sge: Model I -- with two hot stars, forming contact or near-contact configuration 
(Herbig {\it et al.} 1965, Mader and Shafter 1997, Lockley {\it et al.} 1999, 
Smak {\it et al.} 2001) and Model II -- with primary component being the white dwarf 
with an accretion disk (Greiner and van Teeseling 1998, Patterson {\it et al.} 1998, 
Hachisu and Kato 2003). 
In both models the more massive secondary component is a main sequence star 
transfering mass to the primary. 

\subsection { Model I }

Our model (Smak {\it et al.} 2001), based on a simple analysis of spectroscopic 
(radial velocity) and photometric data, resulted in determination of system 
parameters. Its hypothetical part dealt with the hot, expanding gaseous envelope 
(or stellar wind), being the source of the ultraviolet and soft X-ray radiation. 
This requires further support. 

Herbig {\it et al.} (1965) were the first to note that the emission spetrum of V Sge 
is similar to that of Wolf-Rayet stars. In what follows we will show that the 
same is true with respect to the mass outflow rate, the X-ray luminosity and 
the bolometric luminosity of the primary component. 

As discussed in Section 3, the rate of mass loss from the system  
was obtained by Lockley {\it et al.} (1997,1999) from radio observations. 
We adopted (cf. Eq.8): $\log \dot M_{1+2}=-4.6\pm 0.3$. 
Both radio observations were made when the observed brightness of V Sge was 
$V=10.9\pm 0.2$ (cf. Lockley {\it et al.} 1999, Fig.9); the corresponding 
luminosity of the primary component (see Appendix) was $\log L_1=39.1\pm 0.3$. 
The X-ray luminosity, as obtained by Eracleous {\it et al.} (1991) at an assumed 
distance of 2.7 kpc, was $L_x=3^{+2}_{-1}\times 10^{32}$. Using $d=4$ kpc we adopt 
$\log L_x=32.8\pm 0.2$. 
At the time when the X-ray flux was measured the observed brightness of V Sge 
was $V=11.0\pm 0.1$ (\v{S}imon and Mattei 1999, Fig.3); the corresponding 
luminosity of the primary component (see Appendix) was $\log L_1=39.0\pm 0.2$. 

We now compare V Sge with binary Wolf-Rayet stars using data from Tables 1 and 3 
of Naz\'e {\it et al.} (2021). 
Results, presented in Fig.3, show that -- in all three plots -- 
V Sge is indistinguishable from binary Wolf-Rayet stars. This implies that -- like 
in those binaries -- the primary component must be a hot, evolved star loosing 
its mass.  

\begin{figure}[htb]
\vskip -5truemm
\epsfysize=11.0cm 
\hspace{2.0cm}
\epsfbox{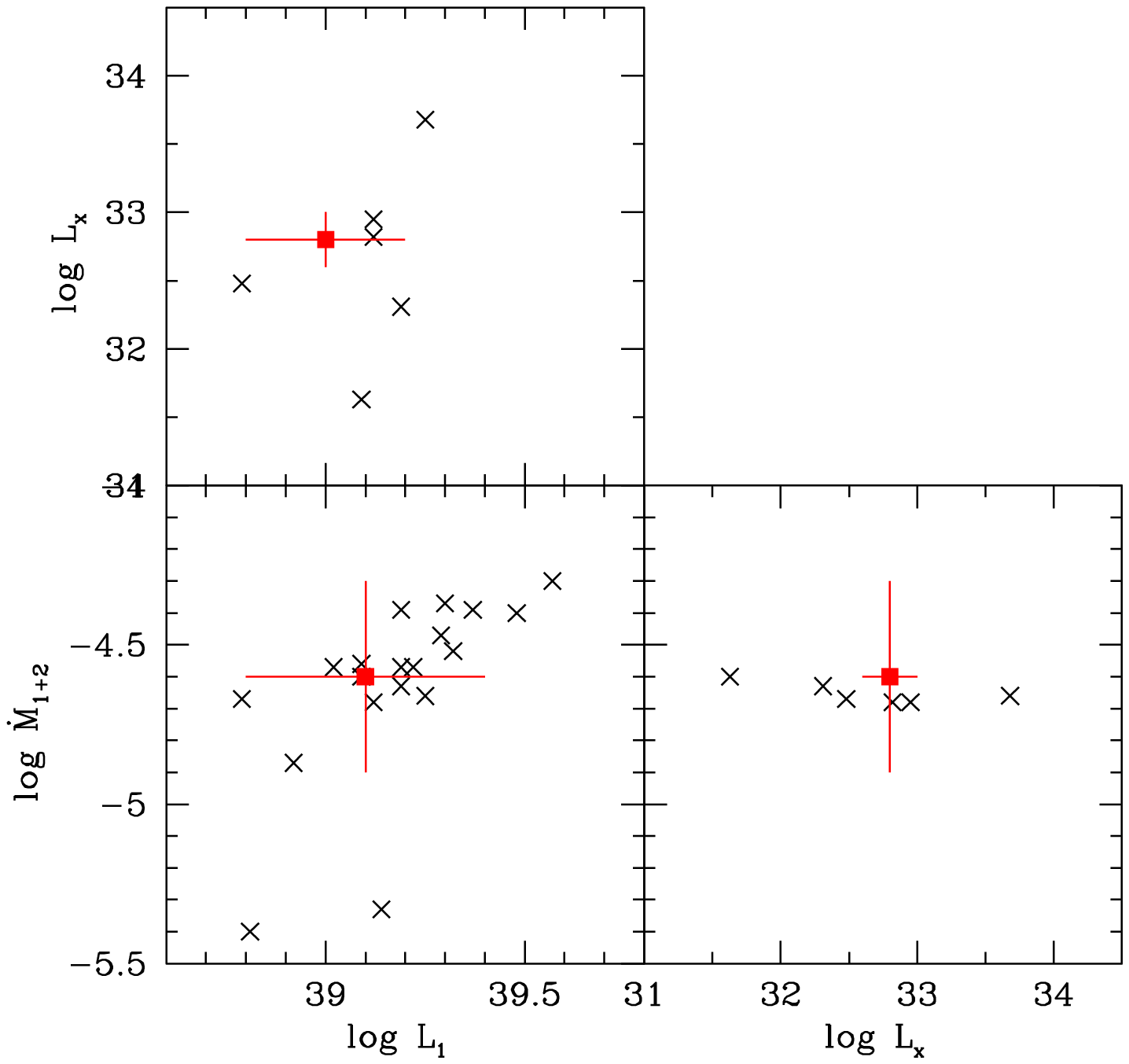} 
\vskip -25truemm
\FigCap { Comparison of V Sge ({\it red symbols}) with binary Wolf-Rayet stars 
involving the luminosity of the primary component $L_1$, the rate of mass outflow 
from the system $\dot M_{1+2}$ and the X-ray luminosity $L_x$. 
}
\end{figure}

\subsection { Model II } 

This model fails to explain the following facts: 

(1) The accretion rate needed to explain the large depth of the primary 
eclipse in the faint state would be very high and -- consequently -- the secondary 
should be strongly heated by the boundary layer. No significant irradiation effect, 
however, is seen around the secondary eclipse (cf. Fig.5 in Smak {\it et al.} 2001). 

(2) In this model the soft X-ray flux must come from the boundary layer between 
the disk and the white dwarf. It is obvious that the X-ray light curve should be 
dominated by a deep primary eclipse. Hoard {\it et al.} (1996) found however that 
the X-ray flux from V Sge, measured by ROSAT, is modulated with periods equal to 
1/2, 1/3, and 1/4 of the orbital period, but {\it not} with the orbital period itself. 

(3) The double fluorescent OIII emission lines are narrow and define regular, 
sinusoidal radial velocity curves with nearly identical values of $\gamma_1$ and 
$\gamma_2$ (Herbig {\it et al.} 1965). This implies that they must come from 
the surfaces of the two components. 
The hypothetical white dwarf would be totally eclipsed between phases $\phi=0.94$ 
and 0.06. Consequently its OIII component should also be eclipsed. 
Noting that around phase zero the two OIII components cannot be separated 
we analyze their combined equivalent widths $W_{0.65}(1+2)$ (in units of continuum 
at phase $\phi=0.65$), listed in Herbig {\it et al.} (1965, Table III).
Their mean intensity outside the primary eclipse is $<W_{0.65}(1+2)>=2.20\pm 0.08$. 
The mean intensity of component 1 outside eclipse is 
$<W_{0.65}(1)>=0.70\pm 0.04$. Should this component be eclipsed we would 
expect -- during eclipse -- $W_{0.65}(1+2)=1.50\pm 0.09$. 
Instead the observed mean value is $<W_{0.65}(1+2)>=2.21\pm 0.19$,  
which is practically the same as outside eclipse. 

(4) From our analysis of the orbital period variations it follows that the primary 
component is loosing mass at a rate $\dot M_1=-4\times 10^{-7}M{\odot}/yr$ 
or -- possibly -- much higher. Such a significant mass loss from the white dwarf 
is practically impossible. 

Hachisu and Kato (2003) made an attempt to overcome some of those problems  
by presenting a model based on a number of arbitrary, unrealistic assumptions and involving several extra free parameters. 
In their model the accretion disk (see their Figs 3, 4 and 6) not only has a very 
peculiar shape of the outer edge and extends beyond its tidal radius, but penetrates 
through the secondary component(!)... Such a model cannot be taken seriously.

\section { Discussion } 

One of the results presented above, namely that concerning the accelerated 
mass tranfer from the more massive secondary component (Section 3), requires 
further comments. 
Situation here is similar to that during the first stage of the 
evolution of a close binary leading to the Algol phase (cf. Paczy\'nski 1971): 
the process of mass exchange is unstable and occurs on a thermal time scale. 
It is obvious that the mass tranfer rate must initially increase 
(up to a certain maximum) and then decrease. 
The observed acceleration of the mass transfer in V Sge is therefore not unusual.  

\section { Appendix: Luminosities of the Primary Component } 

To estimate the bolometric luminosity of the primary component at various 
levels of brightness we use the data and results contained in 
Tables 1 and 2 of Smak {\it et al.} (2001). Results corresponding to the 
black body and Kurucz fluxes are presented in Table 2.

\begin{table}[h!]
{\parskip=0truept
\baselineskip=0pt {
\medskip
\centerline{Table 2}
\medskip
\centerline{ Luminosities of the Primary Component }
\medskip
$$\offinterlineskip \tabskip=0pt
\vbox {\halign {\strut
\vrule width 0.5truemm #&	
\quad#\quad\hfil&            
\vrule#&			
\hfil\quad#\quad\hfil&            
\vrule#&			
\hfil\quad#\quad\hfil&            
\vrule#&			
\hfil\quad#\quad\hfil&            
\vrule width 0.5 truemm # \cr	
\noalign {\hrule height 0.5truemm}
&Description&& V &&$\log L_1^{BB}$&&$\log L_1^{Ku}$ &\cr
\noalign {\hrule height 0.5truemm}
&HPSP-faint1&& 12.51 && 37.96 && 38.16 &\cr
&HPSP-faint2&& 12.12 && 38.21 && 38.38 &\cr
&MS-low     && 11.70 && 38.18 && 38.32 &\cr
&HPSP-bright&& 10.87 && 39.12 && 39.28 &\cr
&MS-high    && 10.76 && 39.25 && 39.34 &\cr
\noalign {\hrule height 0.5truemm}
}}$$
}}
\end{table}

Formal fit to the 
points gives the following relation between the luminosity of the primary 
component and the observed brightness of the system 

\beq
\dot \log L_1~=~38.323(0.068)~-~0.727(0.086)\times (V-12)~. 
\eeq

\begin {references} 


\refitem {Eracleous, M., Halpern, J. and Patterson, J.} {1991} {\ApJ} {382} {290}

\refitem {Greiner, J. and van Teeseling, A.} {1998} {\AA} {339} {L21}

\refitem {Hachisu, I. and Kato, M.} {2003} {\ApJ} {598} {527}

\refitem {Herbig, G.H., Preston, G.W., Smak, J. and Paczy\'nski, B.} {1965} 
         {\ApJ} {141} {617} 

\refitem {Hoard, D.W., Wallerstein, G. and Willson, L.A.} {1996} {\PASP} {108} {81}

\refitem {H{\"o}bscher, J.} {2005} {\it IBVS} {~} {5643}

\refitem {H{\"o}bscher, J.} {2016} {\it IBVS} {~} {6157} 

\refitem {H{\"o}bscher, J. and Lehmann, P.B.} {2015} {\it IBVS} {~} {6149} 

\refitem {H{\"o}bscher, J., Lehmann, P.B., Monninger, G. and  
    Steinbech, H.- F.} {2010} {\it IBVS} {~} {5941} 

\refitem {H{\"o}bscher, J., Steinbech, H.-M. and Walter, F.} {2008} {\it IBVS} 
    {~} {5830} 

\refitem {Koch, R.H., Corcoran, F.M., Holenstein, B.D. and McCluskey, G.E.} {1986} 
    {\ApJ} {306} {618}

\refitem {Lockley, J.J., Eyres, S.P.S. and Wood, J.H. }
   {1997} {\MNRAS} {287} {L14}

\refitem {Lockley, J.J., Wood, J.H., Eyres, S.P.S., Naylor, T. and Shugarov, S.}
   {1999} {\MNRAS} {310} {963}

\refitem {Mader, J.A. and Shafter, A.W.} {1997} {\PASP} {109} {1351} 

\refitem {Naz\'e, Y., Gosset, E. and Quentin, M.} {2021} {\MNRAS} {501} {4214}

\refitem {Ogloza, W., Dr\'{o}\.{z}d\.{z}, M. and Zo{\l}a, S.} {2000} 
         {\it IBVS} {~} {4847} 

\refitem {Paczy\'nski, B.} {1971} {Ann.Rev.Astron.Astrophys.} {9} {183}

\refitem {Parimucha, S. et al.} {2007} {\it IBVS} {~} {5777} 

\refitem {Patterson, J. et al.} {1998} {\PASP} {110} {380}

\refitem {Pribulla, T. et al.} {2005} {\it IBVS} {~} {5668}

\refitem {\v{S}imon, V. and Mattei, J.A.} {1999} {Astron.Astrophys.Suppl.Ser.} 
     {139} {75}

\refitem {Smak, J.} {1967} {\Acta} {17} {55} 

\refitem {Smak, J.I., Belczy\'nski, K. and Zo{\l}a, S.} {2001} {\Acta} {51} {117}

\refitem {Zejda, M.} {2002} {\it IBVS} {~} {5287}

\end {references}

\end{document}